\documentclass[twocolumn,showpacs,preprintnumbers,amsmath,amssymb,letter]{revtex4}
\usepackage{graphicx}
\usepackage{dcolumn}
\usepackage{bm}
\usepackage{epsfig}
\usepackage{slashed}
\usepackage{simplewick}

\newcommand{\half}{\frac{1}{2}}
\newcommand\beq{\begin{eqnarray}}
\newcommand\eeq{\end{eqnarray}} 
\newcommand\eqn[1]{\label{eq:#1}} 
\newcommand\eq[1]{Eq.~(\ref{eq:#1})} 
\newcommand{\vev}[1]{\langle #1 \rangle}

\newcommand{\bfA}{{\mathbf A}}
\newcommand{\bfa}{{\mathbf a}}

\newcommand{\bfv}{{\mathbf v}}

\newcommand{\CB}{{\cal B}}

\newcommand{\CD}{{\cal D}}

\newcommand{\CO}{{\cal O}}

\newcommand{\CN}{{\cal N}}

\newcommand{\Tr}{{\rm Tr\,}}

\newcommand{\mybar}[1]%
        {\kern  .6pt\overline{\kern - .6pt#1\kern - .6pt}\kern  .6pt}
\newcommand{\mqcd}{\text{QCD}}
\newcommand{\xqcd}{XQCD }
\newcommand{\mxqcd}{\text{XQCD}}
\newcommand{\sym}{S_\text{YM}}

\begin{document}

\preprint{INT-PUB-13-023}

\title{Extended QCD}
 
\author{David B. Kaplan}
\email{dbkaplan@uw.edu}
\affiliation{Institute for Nuclear Theory, Box 351550, Seattle, WA 98195-1550, USA}

 \date{\today}
 
 \begin{abstract}
 I consider a reformulation of QCD which incorporates additional bosonic fields, and analyze it in the large-$N_c$ limit.  This construction makes a more direct connection with the nonrelativistic quark and the chiral quark models, as well as the bag model for baryons.  This extension of QCD is neither a model nor an effective low energy theory, but is exactly equivalent to QCD. 
   \end{abstract}

\maketitle

\section{Introduction}

Two milestones on the path to  the theory of QCD were the invention of the quark model \cite{GellMann:1964nj,zweig1964model}, and the discovery that pions arise as Goldstone bosons due to a chiral symmetry breaking condensate of fermions in the vacuum \cite{Nambu:1961tp}.  The latter phenomenon has been verified in lattice QCD  (where the vacuum exhibits a quark condensate, rather than the nucleon condensate originally envisaged), but the quark model remains just a heuristic picture -- albeit a  successful one -- with no direct connection to QCD beyond sharing fermion degrees of freedom and some of the same symmetries.  In fact, the quark model, with its massive, weakly interacting quarks, has no realization of chiral symmetry, and no way to explain a massless pion in the chiral limit.  An attempt to reconcile the quark model with spontaneously broken chiral symmetry was made in the chiral quark model of Georgi and Manohar \cite{Manohar:1983md} where quarks and gluons coexist with pions at low energy. The model assumes a weak gluon coupling below the chiral symmetry breaking scale in order to explain why constituent quarks appear to be weakly interacting, but results in  an unrealistically low confinement scale and light glueballs.  The model also does not quantitatively explain what happens to the quark-antiquark bound state with the quantum numbers of the pion, and similarly has not been derived from QCD.  Recently the energy range over which this model should apply was extended by invoking the large-$N_c$ expansion \cite{Weinberg:2010bq}, but the basic question of how weakly interacting (yet confined) massive quarks  can arise from QCD remains unresolved.

In this Letter I show how one can extend QCD to contain additional bosonic fields in such a way that makes both the weakly interacting massive quarks of the quark model and  Goldstone boson nature of the pions evident features of the theory.  This version of QCD is neither a model nor a low energy effective theory, but is simply a reformulation of  QCD  with additional redundant variables ---  referred to as Extended QCD, or XQCD --- and which has exactly the same matrix elements  as QCD at all energies.    After showing how to derive this theory, I investigate its properties  in the large-$N_c$ limit, where it is particularly simple to understand, and show that the quark model picture emerges of weakly interacting fermions with constituent masses for mesons other than the lightest $0^\pm$ states.  In contrast,  baryons in the large-$N_c$ limit of XQCD appear to be  bag-like objects, perhaps somewhat similar to Skyrmions or the chiral bag model representation.  In the discussion section I suggest various possible applications of XQCD, and offer a speculation about confinement.

\section{Defining XQCD}

For the rest of this Letter the term ``QCD" refers to a generalization of real QCD that has $N_f$ light flavors of quarks and gauge group $SU(N_c)$; for simplicity  only the case of degenerate quarks is considered.  The theory is assumed to be formulated in Euclidian spacetime and regulated in a gauge invariant way that preserves chiral symmetry, such as on a lattice with domain wall \cite{Kaplan:1992bt}  or overlap fermions \cite{Neuberger:1997fp}.

It is simple to add a scalar field to QCD without changing physics:  for example, one can insert into the QCD path integral the gaussian path integral 
\beq
\int [d\sigma]\, e^{-\int d^4x\, \lambda^2(\sigma - \mybar \psi \psi)^2}
\eqn{sig}
\eeq
where $\psi$ are the quark fields and $\sigma$ is a new scalar field; this integral is just a constant, which can be normalized to unity.  Expanding the exponent in \eq{sig}, the new theory then looks like QCD with a massive, non-propagating $\sigma$ field with a Yukawa interaction to the quarks, plus a compensating repulsive four-quark contact interaction.  This particular example leads to  chiral symmetry breaking interactions, but it is possible to replace $\sigma$ with a field $\Phi$ transforming as a  $(N,\mybar N)$ bifundamental under the $SU(N_f)\times SU(N_f)$ chiral symmetry with  symmetric couplings to quarks and kinetic term $\Tr\Phi^\dagger\Phi$ \footnote{The $\Phi$ field is in general complex, but in the special case of $N_f=2$ on can impose the reality condition $\Phi = \sigma_2\Phi^*\sigma_2$, in which case a factor of $1/2$ should precede the  $\Tr\Phi^\dagger\Phi$ operator.}.  Adding an auxiliary  field to QCD might look unnecessarily complicated  and at the same time trivial; that it is obvious this procedure does not alter QCD is a virtue, while it should be noted that correlations of the $\Phi$ field are not trivial at all,  exhibiting all the nonanalytic structure associated with the $0^+$ meson sector of QCD and  a  vacuum expectation value induced by the quark condensate.   The benefit of adding such a field, however,  is   obscured by the compensating four-fermion contact interaction.  This can be eliminated by introducing new, colored auxiliary  fields,  flavor singlet vector and axial vector fields $\bfv_\mu$ and $\bfa_\mu$ which are hermitian $N_c\times N_c$ matrices, transforming as a singlet plus adjoint under the $SU(N_c)$ gauge group \footnote{Boldface indicates  two-index tensors under $U(N_c)$.} by making use of the Fierz relation
\begin{equation}
\begin{aligned}
(P_+)_{ij}(P_-)_{kl} &+ (P_-)_{ij}(P_+)_{kl} \cr & =\frac{1}{4}\left[(\gamma_\mu)_{il}(\gamma_\mu)_{kj} - (\gamma_\mu\gamma_5)_{il}(\gamma_\mu\gamma_5)_{kj}\right]
\eqn{Fierznf}
\end{aligned}
\end{equation}
where $P_\pm = \half(1\pm\gamma_5)$. This yields the  path integral identity
\beq
\int 
e^{-N_c\lambda^2\int d^4 x\, \left[\Tr \Phi^\dagger\Phi +\half \Tr(\bfv_\mu \bfv_\mu + \bfa_\mu \bfa_\mu)\right]}\,\det\left[\CD + m\right] \cr  = \CN\det\left[\slashed{D} +m\right]\ ,
\eqn{detid}\eeq
where I define
\beq
\CD = \slashed{D}  + \slashed{\bfv} +i \slashed{\bfa} \gamma_5 + 2\left( \Phi P_+ + \Phi^\dagger P_- \right)\ ,
\eqn{cddef}
\eeq
where $D_\mu = (\partial_\mu + i \bfA_\mu)$ is the $SU(N_c)$ covariant derivative, $\bfA_\mu$ being the gluon field. The above path integration is over $\Phi$, $\bfv_\mu$ and $\bfa_\mu$,  and the proportionality constant $\CN$ depends on the mass parameter $\lambda$, but is independent of $\bfA_\mu$.  Note that the relative $i$ between the $\bfv_\mu$ and $\bfA_\mu$ couplings implies that $\bfv_\mu$ exchange is repulsive where gluon exchange is attractive.  When $\bfa_\mu$ exchange is taken into account as well, the repulsion is specifically between right-handed and left-handed flavor-singlet colored quark currents. The introduction of the chiral mean field $\Phi$ is concomitant with weakening of the interaction between quarks.

With this identity the XQCD  action can be defined as
\begin{equation}
\begin{aligned}
&S_\mxqcd=N_c\int d^4x\,\Bigl[\mybar\psi\left( \CD +m\right)\psi
\cr&\quad\ +
\sym +\lambda^2 \Bigl(\Tr\Phi^\dagger\Phi + \half\Tr[\bfv^2_\mu+\bfa^2_\mu ]\Bigr)\Bigr]\ .
\end{aligned}
\end{equation}
where $\sym$ is the conventional Yang-Mills action given the above normalization for the gauge field. Matrix elements of operators in \xqcd are defined by integration over the gauge, fermion, and auxiliary fields with weight $\exp(-S_\mxqcd)$.  The identity \eq{detid} implies that for an operator $\CO$ involving just gauge or quark fields, $\vev{\CO}_\mxqcd = \vev{\CO}_\mqcd$.
 Since the auxiliary field integration in \xqcd is gaussian,  when auxiliary fields appear in $\CO$ its matrix element in XQCD can be related to an operator matrix element in QCD which, up to computable contact terms, is given by the substitution of the appropriate quark bilinear: $\lambda^2\Phi^i_j\to - (\mybar \psi_{Rj} \psi_L^i)$, $\lambda^2(\bfv_\mu)^a_b\to -\mybar \psi_b\gamma_\mu \psi^a$, 
$\lambda^2(\bfa_\mu)^a_b\to -\mybar \psi_b\gamma_\mu \gamma_5\psi^a$, 
with  $i,j$ and $a,b$ being flavor and color indices respectively.  The vev of $\Phi$ is of particular interest, giving rise to an additional mass contribution $M$ for the quarks,
\beq
\lambda^2 \vev{\Phi_{ij}}_\mxqcd 
= \delta_{ij}\Sigma\ ,\quad M = 2\Sigma/\lambda^2
\eqn{sigvevs}\eeq
 $\Sigma=-\vev{\mybar \psi\psi}_\mqcd$ being the condensate of a single  quark flavor in QCD (normalized to be independent of $N_c$). 
  
\section{\xqcd at Large-$N_c$}

\begin{figure}[t]
\centerline{\includegraphics[width=3.4in]{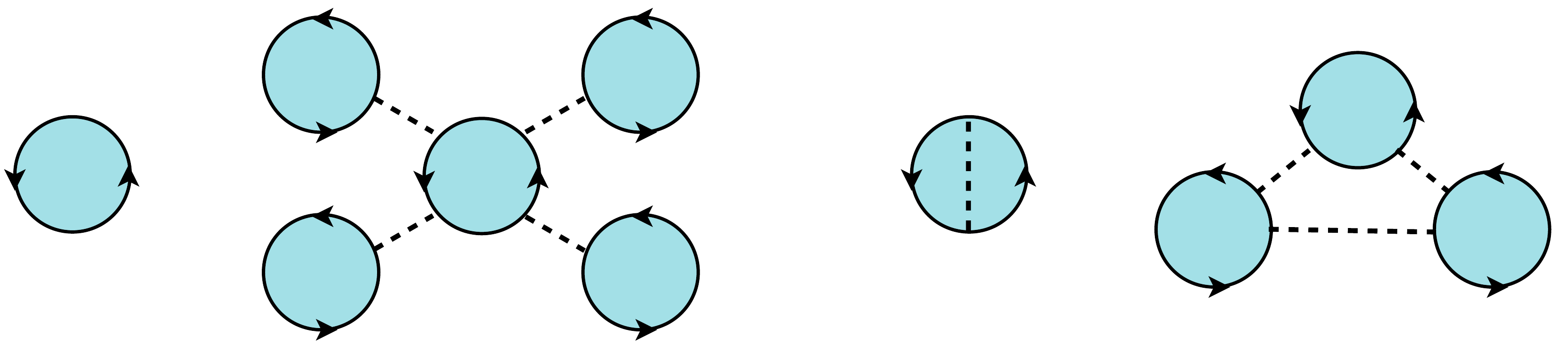}}
\caption{\it Graphs in large-$N_c$ XQCD: quark and $\Phi$ propagators are represented by solid and dashed lines respectively; shaded regions are the sum of all connected planar diagrams involving $\bfA_\mu$, $\bfv_\mu$ and $\bfa_\mu$  fields.  The two diagrams on the left are both $O(N_c)$, while the two on the right are subleading at $O(1)$.}
\label{fig:Ncounting}
\end{figure}

XQCD is most easily understood in the large-$N_c$ limit, which can be described graphically with double color lines for $\bfA_\mu$, $\bfv_\mu$ and $\bfa_\mu$ and a single color line for $\psi$.  Every closed color loop then contributes a factor of $N_c$, as does every vertex, while propagators cost a factor of $1/N_c$ \cite{'tHooft:1973jz}.  Thus every connected planar diagram involving the colored boson fields with a fermion loop as its boundary is order $N_c$, as in QCD, except the graphs now involve $\bfv_\mu$ and $\bfa_\mu$ propagators as well as gluons.     Where XQCD differs significantly from QCD is that any number of such planar diagrams can be connected by $\Phi$ propagators in a tree pattern, and the resultant diagram is still $O(N_c)$; however, the insertion of a $\Phi$ propagator within one of these planar diagrams decreases the order by $1/N_c$ and is therefore subleading, as is any diagram that does not fall apart when cutting a $\Phi$ propagator (see Fig.~\ref{fig:Ncounting}).

The fact that $\Phi$ only contributes through tree diagrams -- where the vertices are themselves are the planar diagrams appearing in Fig.~\ref{fig:Ncounting} -- means that $\Phi$ satisfies classical equations of motion given by
\beq
\frac{\delta\ }{\delta\Phi (x)} \,\int d^4x\,\left( N_c \lambda^2\Tr\Phi^\dagger\Phi - \ln\vev{
\det
(\CD + m)}_\Phi\right)=0
\eqn{eom}\eeq
where $\CD$ is given in \eq{cddef}, and the $\vev{\ldots }_\Phi$ signifies path integration over $\bfA$, $\bfv$ and $\bfa$ with the appropriate action for each field, with a background $\Phi$ field.  
Both terms are proportional to $N_c$ and \eq{eom} is just the saddle point condition, valid at large $N_c$.  This is none other than the gap equation with solution given in \eq{sigvevs}, where we can use our nonperturbative knowledge of QCD to supply the value of the quark condensate.

\subsection{Meson correlation functions}
%
\begin{figure}[t]
\centerline{\includegraphics[height=.76in]{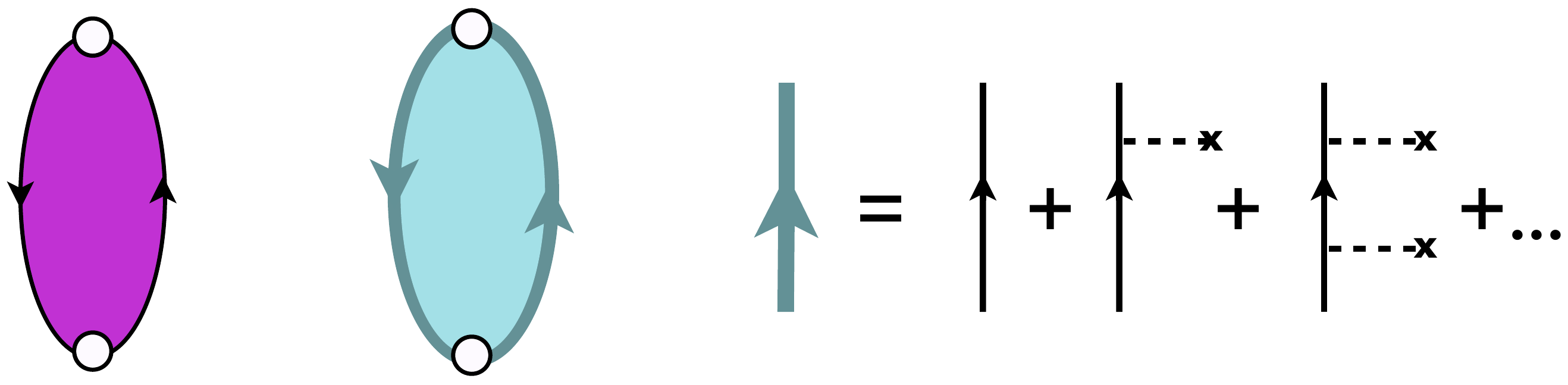}}
\caption{\it The correlator $\vev{\varphi_\Gamma^\dagger(x) \varphi_\Gamma(y)}$ at $O(N_c)$.  Left: the QCD version, the shaded region being the sum of planar gluon contributions.  Right: the XQCD correlator for the case where $\varphi_\Gamma$ cannot mix with $\Phi$, the lightly shaded region representing the sum of planar gluon, $\bfv_\mu$ and $\bfa_\mu$ diagrams, where the fermion propagator is evaluated at $\vev{\Phi} = \Sigma/\lambda^2$.}
\label{fig:rhomeson}
\end{figure}
%
%
%
%
Now consider matrix elements of a meson operator $\varphi_\Gamma = \mybar\psi\Gamma\psi$, where $\Gamma$ is matrix with Dirac and flavor indices.  In QCD the correlation function $\vev{\varphi_\Gamma^\dagger(x) \varphi_\Gamma(y)}_\mqcd$ is given by the sum of all connected planar diagrams of gluons attached to a quark loop with $\varphi_\Gamma$ insertions at $x$ and $y$, as shown on the left in Fig.~\ref{fig:rhomeson}; its Fourier transform is given by a sum of poles corresponding to the tower of meson states with quantum numbers specified by $\Gamma$, all of which are stable in the large-$N_c$ limit.

In \xqcd   $\vev{\varphi_\Gamma^\dagger(x) \varphi_\Gamma(y)}_\mxqcd$  has a different diagrammatic expansion: quark loops are spanned by planar contributions of not only gluons, but also $\bfv$ and $\bfa$ fields, while quark lines are dressed by $\Phi$ trees, such as in Fig.~\ref{fig:Ncounting}, which is equivalent to  giving the quarks the constituent mass $M$.  If $\Gamma$ is chosen so that $\varphi_\Gamma$ has the same quantum numbers as $\Phi$ (a $ 0^\pm$ meson which is either and adjoint of singlet under the vector $SU(N_f)$ flavor symmetry), then there are  additional annihilation diagrams at leading order in $N_c$, as shown in  Fig.~\ref{fig:pimeson}.  In particular, such annihilation diagrams are important for the pion propagator in XQCD.

\subsection{Fixing the scale $\lambda$}
\begin{figure}[t]
\centerline{\includegraphics[height=.76in]{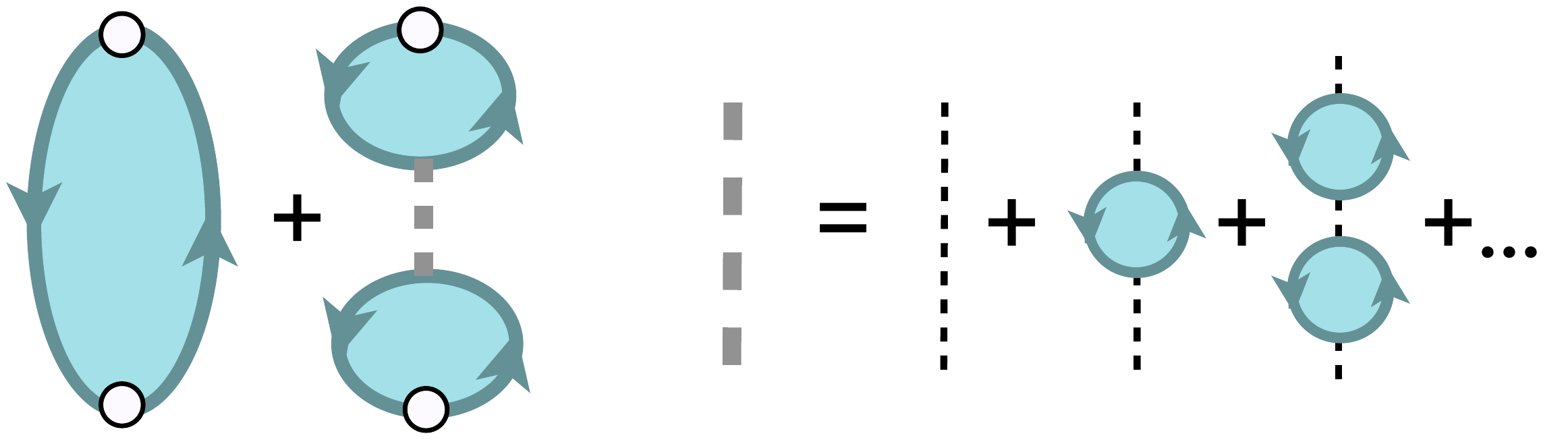}}
\caption{\it The meson correlator in XQCD at $O(N_c)$  for the case where $\varphi_\Gamma$ can mix with $\Phi$, receiving contributions from annihilation diagrams with an intermediary dressed $\Phi$ propagator.}
\label{fig:pimeson}
\end{figure}

A constituent mass $M=2\Sigma/\lambda^2$  is meaningless without specifying the scale $\lambda$, and in general the quarks have strong interactions that modify their mass.  However, it is interesting that if one chooses $\lambda$ such that $M$ is a little more than half the $\rho$ mass, for example, the contribution to the $\rho$ meson correlator is well approximated near the $\rho$ pole by the propagation of free, slightly off-shell quarks in the background $\Phi$ field.   It follows that the sum of nontrivial planar diagrams involving the gluon and  the $\bfv$ and $\bfa$ fields must sum close to zero for this value of $\lambda$ near the $\rho$ pole, meaning that repulsive interactions of the auxiliary fields have nearly cancelled the gluon contribution to the binding of the quarks, and that much of the effects of the gluon in QCD have been subsumed into the constituent mass $M$ in XQCD.  (Not all, however, as confinement implies the absence of a cut in the correlator corresponding to two on-shell quarks). Choosing $\lambda$ such that $M\gtrsim m_\rho/2$ is therefore a reasonable scale setting procedure,  although it can presumably be improved upon by setting $M$ equal to the best fit constituent quark mass in the quark model appropriate for the particular choice of $N_f$, $N_c$ and $m$.  QCD phenomenology and the successes of the nonrelativistic quark model imply that with this scale setting, the strong gluon contributions to the planar diagrams with other flavor and spin quantum numbers will similarly mostly cancel against the repulsive auxiliary fields in all channels (with the exception of the lightest $0^\pm$ mesons) as the nearly free propagation of quarks with mass $M$ correctly accounts for the correlation functions.  In the case of the light pseudoscalars,  the valence contribution to  the correlation function in Fig.~\ref{fig:pimeson} will only have a pole at the heavier $\pi^*$ mass, if the nontrivial planar diagrams all cancel between them, but  the annihilation contribution is guaranteed to have a pole at $p^2=0$ in the chiral limit, as $\Phi$ couples to the Goldstone mode.  In this way the existence of the massless pion is reconciled with the quark model picture of weakly interacting fermions with a substantial constituent mass.

\subsection{Baryon correlation functions}
\begin{figure}[t]
\centerline{\includegraphics[height=.9in]{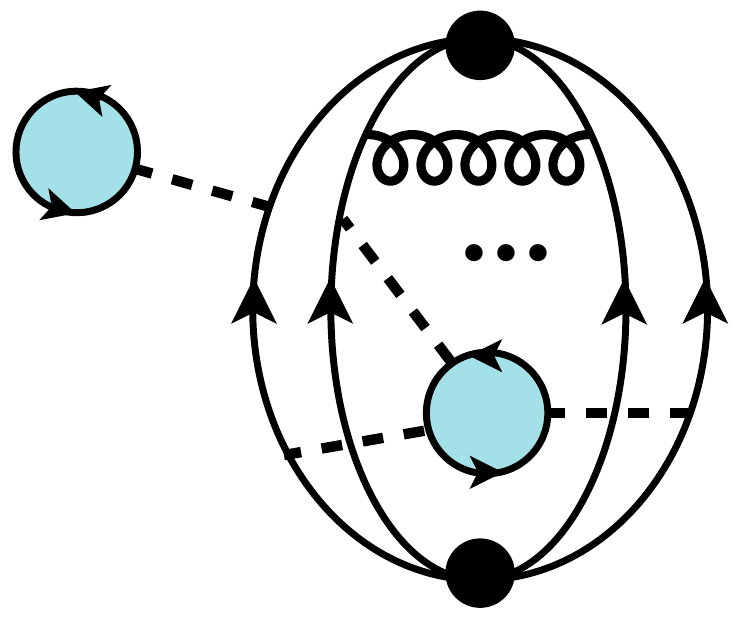}}
\caption{\it A leading order contribution the  baryon correlator in XQCD. }
\label{fig:baryonFig}
\end{figure}

Baryon correlation functions in large-$N_c$ XQCD are intriguingly different than meson correlation functions since one cannot neglect the back reaction of the valence fermions on the saddlepoint solution for $\Phi$. For simplicity, consider the correlation function for  a single baryon, $C_B=\vev{\CB^\dagger(x) \CB(y)}$, with interpolating function
$
\CB(x) = \prod_a \psi_a(x)
$,
where $a$ is the color index, and total antisymmetrization in flavor and Dirac indices is implicit.  Then we expect $C_B \propto \exp(-M_B|x-y|)$ for large $|x-y|$, where $M_B$ is the lightest baryon in the particular quantum number channel selected. As noted by Witten \cite{Witten:1979kh}, gluon (and therefore also $\bfv_\mu$ and $\bfa_\mu$ exchange) between two valence quarks contribute a factor of $O(1/N_c)$, but there are $O(N_c^2)$ such contractions, so the effect is $O(N_c)$, which should be interpreted as an $O(g^2)$ correction to $M_B$.  This implies that in XQCD one must sum up all planar diagrams with exchange of gluons, $\bfv$ and $\bfa$ between the $N_c$ valence quark lines in $C_B$.  Diagrams with $\Phi$ fields receive interesting combinatoric enhancements as well.  As was the case for mesons, the $\Phi$ tadpole diagrams occur at leading order; for example, the tadpole in Fig.~\ref{fig:baryonFig} is $O(1)$ when attached to a single valence quark line, but receives a combinatoric enhancement of $N_c$; the three point function in Fig.~\ref{fig:baryonFig} is $O(1/N_c^2)$, but receives a combinatoric enhancement of $N_c^3$.  One can see that any diagram with $k$ $\Phi$ fields attached to a fermion loop will contribute at $O(N_c)$ to the baryon mass, when those $k$ scalar lines are attached to $k$ valence quark lines. 

Thus one cannot simply replace $\Phi\to\vev{\Phi} =\Sigma/\lambda^2$, the saddle point solution to \eq{eom}.  After integration over the fermion fields, the correlator $\vev{\CB^\dagger(x) \CB(y)}_\mxqcd$ equals the the expectation value $\vev{\det G_{ax,by}}$  averaged over all the boson fields, 
where $G = 1/(\CD+m)$ is the quark propagator, up to an unimportant factor of $1/N_c$.  After integrating over the $\bfA$, $\bfv$, and $\bfa$ fields, this exponentiates as $\ln \vev{\det G_{ax,by}}$ which is $O(N_c)$ and hence the same size as the terms in the saddlepoint condition \eq{eom}.  
Therefore in the presence of a baryon, the diagrams involving $\Phi$ exchange  in Fig.~\ref{fig:baryonFig} may still be replaced by quarks propagating in a classical background $\Phi$ field with only $\bfA$, $\bfv$, and $\bfa$ interactions, but now the classical field is the solution to modified equations of motion,
\begin{equation}
\begin{aligned}
\frac{\delta\ }{\delta\Phi (x)} \,\int d^4x\,&\left( N_c \lambda^2\Tr\Phi^\dagger\Phi - \ln\vev{
\det
(\CD + m)}_\Phi\right.\cr&\left.
-\ln\vev{\det G_{ax,by}}_\Phi\right)=0\ .
\eqn{modeom}
\end{aligned}
\end{equation}
The last term depends on the specific form of the baryon correlation function being measured, but should be independent of the exact form of the interpolating field far away from the points $x,y$ where the baryon is created and destroyed.  I do not venture to try to solve this equation here, but it is likely to be a chiral bag-like solution \cite{Chodos:1975ix}, with the valence quark propagation being concentrated in a region where $\vev{\Phi}$ is reduced, while outside this region, $\Phi$ regains its vacuum magnitude.  Presumably in the case of infinite nuclear matter at large-$N_c$ there is a solution similar to the Skyrme crystal found in Ref.~\cite{Klebanov:1985qi}, and perhaps also noncrystalline solutions still at sufficiently large value of $N_c$ to justify the saddlepoint approximation \cite{Shuster:1999tn}.

\section{Future directions}

XQCD seems to give a more natural picture than QCD for how the quark model emerges consistent with chiral symmetry, essentially by taking some of the strength out of the gluon interaction and putting it into a chiral mean field; whether it provides more quantitative benefits remains to be shown. Up to this point I have focused on chiral symmetry breaking effects and have not said much about confinement.  Since XQCD has the same matrix elements as QCD, it is clear that confinement must exist in XQCD, and it is interesting  to speculate that the color adjoint $\bfv$ and $\bfa$ fields could play an active role, by exhibiting magnetic monopole singularities, for example.  Since XQCD is valid at all energies, this framework might also be useful to better understand the transition between the quark and parton model pictures of hadrons at increasing energy.  

Lattice studies of XQCD may prove impractical since the operator $\CD$ in \eq{cddef} has a complex determinant, even at zero baryon chemical potential, presenting an obstacle for Monte Carlo simulation.  This is unfortunate,  the original motivation for this study being the search for a solution to the  QCD sign problem at nonzero baryon density on the lattice, along the lines of  Refs.~\cite{Grabowska:2012ik,Nicholson:2012xt}.

\begin{acknowledgments}

I gratefully acknowledge discussions with T. Cohen, D. Grabowska, A. Li, A. Manohar, A. Nicholson, M.Savage and K. Splittorff.  This work supported in part by U.S.\ DOE grant No.\ DE-FG02-00ER41132.
\end{acknowledgments}

  \bibliography{QCDsign}

\begin{thebibliography}{14}
\expandafter\ifx\csname natexlab\endcsname\relax\def\natexlab#1{#1}\fi
\expandafter\ifx\csname bibnamefont\endcsname\relax
  \def\bibnamefont#1{#1}\fi
\expandafter\ifx\csname bibfnamefont\endcsname\relax
  \def\bibfnamefont#1{#1}\fi
\expandafter\ifx\csname citenamefont\endcsname\relax
  \def\citenamefont#1{#1}\fi
\expandafter\ifx\csname url\endcsname\relax
  \def\url#1{\texttt{#1}}\fi
\expandafter\ifx\csname urlprefix\endcsname\relax\def\urlprefix{URL }\fi
\providecommand{\bibinfo}[2]{#2}
\providecommand{\eprint}[2][]{\url{#2}}

\bibitem[{\citenamefont{Gell-Mann}(1964)}]{GellMann:1964nj}
\bibinfo{author}{\bibfnamefont{M.}~\bibnamefont{Gell-Mann}},
  \bibinfo{journal}{Phys.Lett.} \textbf{\bibinfo{volume}{8}},
  \bibinfo{pages}{214} (\bibinfo{year}{1964}).

\bibitem[{\citenamefont{Zweig and An}(1964)}]{zweig1964model}
\bibinfo{author}{\bibfnamefont{G.}~\bibnamefont{Zweig}} \bibnamefont{and}
  \bibinfo{author}{\bibfnamefont{S.}~\bibnamefont{An}}, \bibinfo{journal}{CERN
  Report} \textbf{\bibinfo{volume}{8419}} (\bibinfo{year}{1964}).

\bibitem[{\citenamefont{Nambu and Jona-Lasinio}(1961)}]{Nambu:1961tp}
\bibinfo{author}{\bibfnamefont{Y.}~\bibnamefont{Nambu}} \bibnamefont{and}
  \bibinfo{author}{\bibfnamefont{G.}~\bibnamefont{Jona-Lasinio}},
  \bibinfo{journal}{Phys.Rev.} \textbf{\bibinfo{volume}{122}},
  \bibinfo{pages}{345} (\bibinfo{year}{1961}).

\bibitem[{\citenamefont{Manohar and Georgi}(1984)}]{Manohar:1983md}
\bibinfo{author}{\bibfnamefont{A.}~\bibnamefont{Manohar}} \bibnamefont{and}
  \bibinfo{author}{\bibfnamefont{H.}~\bibnamefont{Georgi}},
  \bibinfo{journal}{Nucl.Phys.} \textbf{\bibinfo{volume}{B234}},
  \bibinfo{pages}{189} (\bibinfo{year}{1984}).

\bibitem[{\citenamefont{Weinberg}(2010)}]{Weinberg:2010bq}
\bibinfo{author}{\bibfnamefont{S.}~\bibnamefont{Weinberg}},
  \bibinfo{journal}{Phys.Rev.Lett.} \textbf{\bibinfo{volume}{105}},
  \bibinfo{pages}{261601} (\bibinfo{year}{2010}), \eprint{1009.1537}.

\bibitem[{\citenamefont{Kaplan}(1992)}]{Kaplan:1992bt}
\bibinfo{author}{\bibfnamefont{D.~B.} \bibnamefont{Kaplan}},
  \bibinfo{journal}{Phys.Lett.} \textbf{\bibinfo{volume}{B288}},
  \bibinfo{pages}{342} (\bibinfo{year}{1992}), \eprint{hep-lat/9206013}.

\bibitem[{\citenamefont{Neuberger}(1998)}]{Neuberger:1997fp}
\bibinfo{author}{\bibfnamefont{H.}~\bibnamefont{Neuberger}},
  \bibinfo{journal}{Phys.Lett.} \textbf{\bibinfo{volume}{B417}},
  \bibinfo{pages}{141} (\bibinfo{year}{1998}), \eprint{hep-lat/9707022}.

\bibitem[{\citenamefont{'t~Hooft}(1974)}]{'tHooft:1973jz}
\bibinfo{author}{\bibfnamefont{G.}~\bibnamefont{'t~Hooft}},
  \bibinfo{journal}{Nucl.Phys.} \textbf{\bibinfo{volume}{B72}},
  \bibinfo{pages}{461} (\bibinfo{year}{1974}).

\bibitem[{\citenamefont{Witten}(1979)}]{Witten:1979kh}
\bibinfo{author}{\bibfnamefont{E.}~\bibnamefont{Witten}},
  \bibinfo{journal}{Nucl.Phys.} \textbf{\bibinfo{volume}{B160}},
  \bibinfo{pages}{57} (\bibinfo{year}{1979}).

\bibitem[{\citenamefont{Chodos and Thorn}(1975)}]{Chodos:1975ix}
\bibinfo{author}{\bibfnamefont{A.}~\bibnamefont{Chodos}} \bibnamefont{and}
  \bibinfo{author}{\bibfnamefont{C.~B.} \bibnamefont{Thorn}},
  \bibinfo{journal}{Phys.Rev.} \textbf{\bibinfo{volume}{D12}},
  \bibinfo{pages}{2733} (\bibinfo{year}{1975}).

\bibitem[{\citenamefont{Klebanov}(1985)}]{Klebanov:1985qi}
\bibinfo{author}{\bibfnamefont{I.~R.} \bibnamefont{Klebanov}},
  \bibinfo{journal}{Nucl.Phys.} \textbf{\bibinfo{volume}{B262}},
  \bibinfo{pages}{133} (\bibinfo{year}{1985}).

\bibitem[{\citenamefont{Shuster and Son}(2000)}]{Shuster:1999tn}
\bibinfo{author}{\bibfnamefont{E.}~\bibnamefont{Shuster}} \bibnamefont{and}
  \bibinfo{author}{\bibfnamefont{D.}~\bibnamefont{Son}},
  \bibinfo{journal}{Nucl.Phys.} \textbf{\bibinfo{volume}{B573}},
  \bibinfo{pages}{434} (\bibinfo{year}{2000}), \eprint{hep-ph/9905448}.

\bibitem[{\citenamefont{Grabowska et~al.}(2013)\citenamefont{Grabowska, Kaplan,
  and Nicholson}}]{Grabowska:2012ik}
\bibinfo{author}{\bibfnamefont{D.}~\bibnamefont{Grabowska}},
  \bibinfo{author}{\bibfnamefont{D.~B.} \bibnamefont{Kaplan}},
  \bibnamefont{and} \bibinfo{author}{\bibfnamefont{A.~N.}
  \bibnamefont{Nicholson}}, \bibinfo{journal}{Phys.Rev.}
  \textbf{\bibinfo{volume}{D87}}, \bibinfo{pages}{014504}
  (\bibinfo{year}{2013}), \eprint{1208.5760}.

\bibitem[{\citenamefont{Nicholson et~al.}(2013)\citenamefont{Nicholson,
  Grabowska, and Kaplan}}]{Nicholson:2012xt}
\bibinfo{author}{\bibfnamefont{A.~N.} \bibnamefont{Nicholson}},
  \bibinfo{author}{\bibfnamefont{D.}~\bibnamefont{Grabowska}},
  \bibnamefont{and} \bibinfo{author}{\bibfnamefont{D.~B.}
  \bibnamefont{Kaplan}}, \bibinfo{journal}{J.Phys.Conf.Ser.}
  \textbf{\bibinfo{volume}{432}}, \bibinfo{pages}{012032}
  (\bibinfo{year}{2013}), \eprint{1210.7250}.

\end{thebibliography}
\end{document}